\documentclass[twocolumn]{svjour3} 
\usepackage[english]{babel}
\usepackage{amsmath,amssymb}
\usepackage{xspace}
\usepackage{overpic}
\usepackage{upgreek}
\usepackage{hyperref}

\def\nue        {\ensuremath{\nu_e}\xspace}
\def\num        {\ensuremath{\nu_\mu}\xspace}
\def\nut        {\ensuremath{\nu_\tau}\xspace}

\def\epem       {\ensuremath{e^+e^-}\xspace}
\newcommand{\mev}{\ensuremath{\mathrm{\,Me\kern -0.1em V}}\xspace}
\def\epm        {\ensuremath{e^\pm}\xspace} 
\def\mupm        {\ensuremath{\mu^\pm}\xspace} 
\def\taupm       {\ensuremath{\tau^\pm}\xspace} 

\def\taum       {\ensuremath{\tau^-}\xspace}
\def\nueb       {\ensuremath{\nub_e}\xspace}
\def\nub        {\ensuremath{\overline{\nu}}\xspace}
\def\numb       {\ensuremath{\nub_\mu}\xspace}
\def\nutb       {\ensuremath{\nub_\tau}\xspace}
\def\Bbar    {\kern 0.18em\overline{\kern -0.18em B}{}\xspace}
\def\nulb       {\ensuremath{\nub_\ell}\xspace}
\def\Bub     {\ensuremath{B^-}\xspace}

\def\Bzb     {\ensuremath{\Bbar^0}\xspace}

\newcommand{\gev}{\ensuremath{\mathrm{\,Ge\kern -0.1em V}}\xspace}
\def\Bm      {\ensuremath{\Bub}\xspace}
\def\m    {\ensuremath{{\rm \,m}}\xspace}
\def\B       {\ensuremath{B}\xspace}
\def\BB      {\ensuremath{B\Bbar}\xspace} 
\def\BpBm    {\ensuremath{\Bu {\kern -0.16em \Bub}}\xspace}
\def\Bu      {\ensuremath{B^+}\xspace}
\def\BzBzb   {\ensuremath{\Bz {\kern -0.16em \Bzb}}\xspace}
\def\Bz      {\ensuremath{B^0}\xspace}

\def\Bp      {\ensuremath{\Bu}\xspace}
\def\pbar     {\ensuremath{\overline{p}}\xspace}

\newcommand{\Btag}{\ensuremath{B_{\rm tag}}\xspace}
\newcommand{\Bsig}{\ensuremath{B_{\rm sig}}\xspace}

\def\mupm{\ensuremath{\mu^{\pm}}\xspace}

\def\Upsil{\ensuremath{\Upsilon(4S)}\xspace}

\def\BDtaunu{\ensuremath{\Bbar \rightarrow D \tau^-\nutb}\xspace}
\def\BDstaunu{\ensuremath{\Bbar\rightarrow D^* \tau^-\nutb}\xspace}

\def\BDellnu{\ensuremath{\Bbar \rightarrow D \ell^- \nulb}\xspace}
\def\BDsellnu{\ensuremath{\Bbar \rightarrow D^* \ell^- \nulb}\xspace}

\def\BDxtaunu{\ensuremath{\Bbar \rightarrow D^{(*)} \tau^- \nutb}\xspace}
\def\BDxellnu{\ensuremath{\Bbar \rightarrow D^{(*)} \ell^- \nulb}\xspace}

\def\RDs{\ensuremath{{\cal R}_{D^*}}\xspace}
\def\RD{\ensuremath{{\cal R}_{D}}\xspace}
\def\Esl{\ensuremath{E^*_\ell}\xspace}

\def\mmiss{\ensuremath{m_{\rm miss}^2}\xspace}

\title{A Challenge to Lepton Universality in B Meson Decays}
\author{Vera L\"uth}
\institute{
  SLAC National Accelerator Laboratory, Stanford, California 94309, USA\\
  \email{luth@slac.stanford.edu}}

\begin{document}
\flushbottom 
\maketitle 

\begin{abstract}
One of the key assumptions of the Standard Model of fundamental particles is that the interactions of the
charged leptons, namely electrons, muons, and taus, differ {\em only} because of their different masses.
While precision tests 
have not revealed any definite
violation of this assumption, recent studies of $B$ meson decays involving the higher-mass tau lepton have resulted in
observations that challenge lepton universality at the level of four standard deviations. A confirmation of
these results would point to new particles or interactions, and could have profound implications for our
understanding of particle physics.
\end{abstract}


\section*{Motivation}
\label{sec:intro}

More than 70 years of particle physics research have led to an elegant and concise theory of particle interactions at
the sub-nuclear level, commonly referred to as the Standard Model (SM)~\cite{Mann:2010zz,Weinberg:1996kr}.  
%
In the framework of the SM of particle physics the fundamental building blocks, quarks and leptons, are each
grouped in three generations of two members each.  The three charged leptons, the electron ($e^-$), the muon
($\mu^-$) and the tau ($\tau^-$) are each paired with a very low mass, electrically neutral neutrino,
$\nue, \num,$ and $\nut$. The three generations are ordered by the mass $\m_{\ell}$ of the charged lepton ranging 
from 0.511\mev for $\epm$ to
105\mev for $\mupm$, and 1,777\mev for $\taupm$~\cite{Ablikim:2014uzh}.  
Charged leptons participate in electromagnetic and weak, 
whereas neutrinos only undergo weak interaction. 
The SM assumes that these interactions of
the charged and neutral leptons are universal, i.e., the same for the three generations.

Precision tests of lepton universality have been performed by many experiments. 
To date no definite violation of lepton universality has been observed. 
Among the most precise tests is a comparison of decay rates of 
$K$ mesons, $K^- \to e^-
\nueb$ versus $K^-\to \mu^- \numb$~\cite{Lazzeroni:2012cx}~\cite{charge}.
Furthermore, taking into account precision measurements of the tau and muon masses and lifetimes, the measured decay rates $\taum \to e^- \nueb \nut$ and $\mu^- \to e^- \nueb \num$, have confirmed the equality of the weak coupling strengths of the tau and muon~\cite{Ablikim:2014uzh}. %

However, recent studies of semileptonic decays of $B$ mesons of the form $\BDxellnu$, with $\ell =
e, \mu,$ or $\tau$, have resulted in observations that seem to challenge lepton universality.
These weak decays are well understood in the framework
of the SM, and therefore offer a unique opportunity to search for unknown processes, for instance non-SM
couplings to yet undiscovered charged partners of the Higgs boson~\cite{Tanaka:1994ay} or hypothetical lepto-quarks~\cite{Freytsis:2015qca}. Such
searches have been performed on data collected by three different experiments, BABAR and Belle 
at $\epem$ colliders
in the U.S.A. and in Japan, and  LHCb at the proton-proton ($pp$) collider at CERN in Europe.

In the following, details of the measurements, their results and preliminary studies to understand the observed
effects will be presented, along with prospects for improved sensitivity and complementary measurements. This article
is partially based on an earlier review with the same title~\cite{nature}.

\section*{Standard Model Predictions of B Meson Decay Rates}
\label{sec:bdecays}

\begin{figure}[btp!]
\centering
\includegraphics[width=0.34\textwidth]{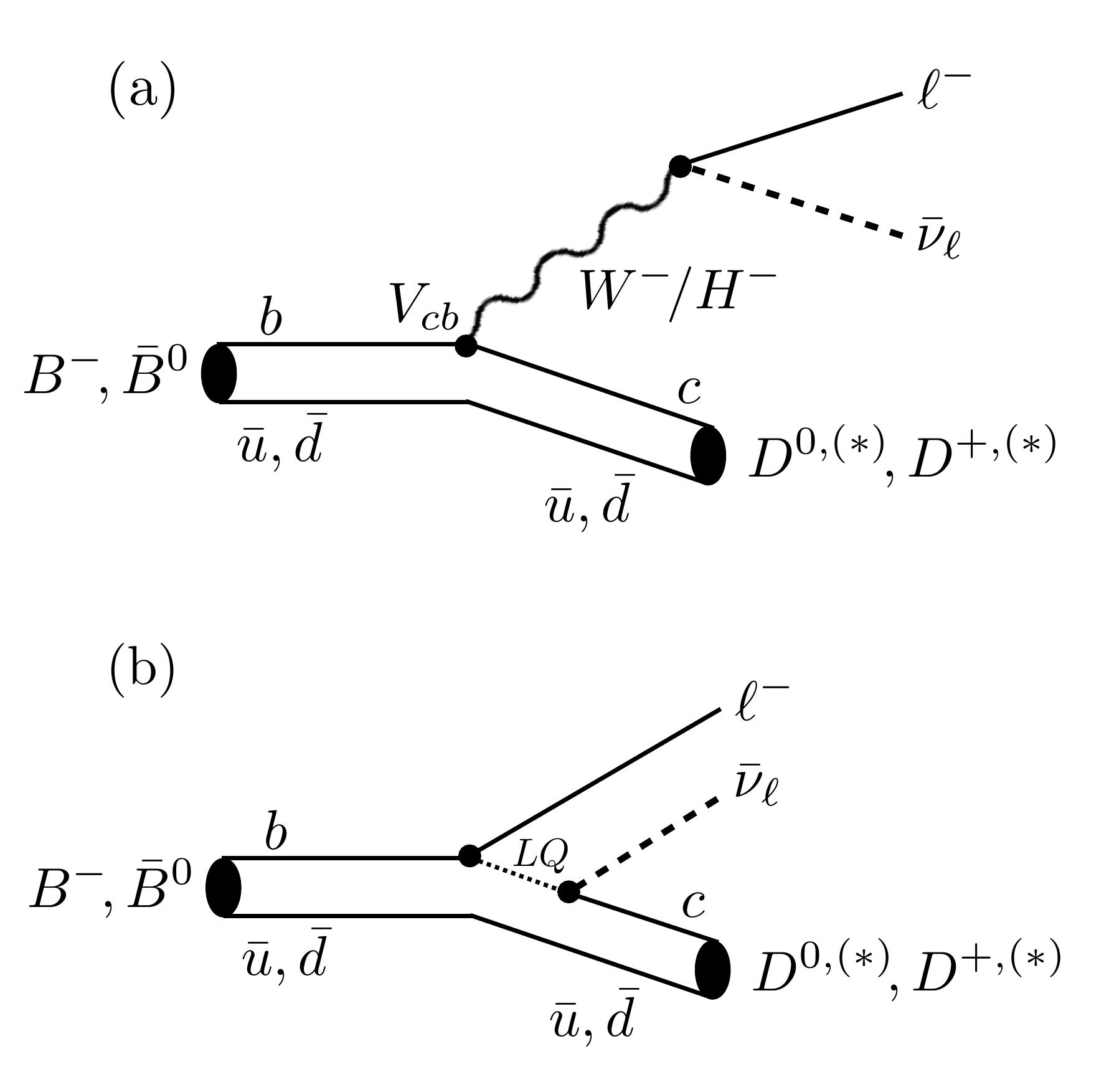}
\caption{Diagrams for decay process for \BDxellnu) decays: 
(a) for a tree level process mediated either by a vector boson $(W^-$)
or a hypothetical spin-0 charged Higgs boson $(H^-)$, or (b) couplings to a hypothetical lepto-quark $(LQ)$. }
\label{fig:feynman}
\end{figure}

According to the SM, semileptonic decays of $B$ mesons are mediated by the $W^-$ boson, as shown schematically in
Figure~\ref{fig:feynman}a.  
The differential decay rate, ${\rm d}\Gamma$, for semileptonic decays involving $D^{(*)}$ mesons depends on both $\m^2_{\ell}$ and $q^2$, the invariant mass squared of the lepton 
pair~\cite{Korner:1989qb},  
\begin{align}
&\frac{{\rm d}\Gamma^{SM}(\BDxellnu)}{{\rm d}q^2}\,   
=  \underbrace{\frac{G_F^2\; |V_{cb}|^2\; |\boldsymbol{p}^*_{D^{(*)}}| \; q^2}{96\pi^3 m_B^2}
  \left(1-\frac{m_\ell^2}{q^2} \right)^2}_{\text{universal and phase space  factors}}
 \\ \nonumber & \times 
\underbrace{\left[(|H_{+}|^2+|H_{-}|^2+|H_{0}|^2) \left(1+\frac{m_\ell^2}{2q^2}  \right) + \frac{3 m_\ell^2}{2q^2}|H_{s}|^2 \right]}_{\text{hadronic effects}}~.
 \label{eq:Gamma_sl}
\end{align}
\noindent
The first factor is universal for all semileptonic $B$ decays, containing a quark flavor mixing parameter~\cite{Kobayashi:1973fv} $|V_{cb}|$~\cite{Amhis:2014hma} for $b \to c$ quark transitions, and $p^*_{D^{(*)}}$,
the 3-momentum of the $D^{(*)}$ meson. The four helicity~\cite{helicity} amplitudes $H_+, H_-, H_0$ and $H_s$ capture the impact of hadronic
effects. They depend on the spin of the charm meson and on $q^2$.  
The much larger $\tau$ mass not only impacts the rate, but also the decay kinematics via the $H_s$ amplitude. All four amplitudes contribute to $\BDsellnu$, 
while only  $H_0$ and $H_s$ contribute to $\BDellnu$, which leads to a higher sensitivity of this decay mode to the scalar contribution $H_s$.
The minimum value of $q^2$ is equal to $m^2_{\ell}$.

Measurements of the ratios of semileptonic branching fractions remove the dependence on $|V_{cb}|$, lead to a partial cancellation of theoretical uncertainties related to hadronic effects, and reduce of the impact of experimental uncertainties. The averages of the
current predictions~\cite{Bigi:2016mdz,Bernlochner:2017jka} and ~\cite{Bernlochner:2017jka,Bigi:2017jbd,Jaiswal:2017rve} for the two ratios are
\begin{eqnarray}
\label{eq:RD}
{\cal R}^{SM}_D     &=& \frac {{\cal B}(\Bbar \to D \tau^- \nutb)}
                              {{\cal B}(\Bbar \to D e^- \nueb)} 
= 0.299 \pm 0.003 \\
\label{eq:RDs}
{\cal R}^{SM}_{D^*} &=&\frac {{\cal B}(\Bbar \to D^* \tau^- \nutb)}
                             {{\cal B}(\Bbar \to D^* e^- \nueb)} 
= 0.258 \pm 0.005 . \
\end{eqnarray}
\noindent
The predicted ratios for ${\cal B}(\Bbar \to D^* \mu^- \numb)$ are identical within the quoted precision.  In the following, $\Bbar \to D^{(*)} \tau^- \nutb)$
decays are referred to as the "signal", and $\Bbar \to D^{(*)} e^- \nueb)$ with 
$\ell = e,\mu$ are referred to as "normalization".

\section*{B Meson Production and Detection}
\label{sec:experiments}

$\B$ meson decays have been studied at 
$pp$ and $e^+e^-$ colliding beam facilities, operating at very different beam energies. 

The $e^+e^-$ colliders operated at a fixed energy of 10.579\gev in the years 1999 to 2010.  At this energy, about 20 \mev above the kinematic threshold for $\BB$ production, $e^+$ and $e^-$ annihilate and produce a particle, 
commonly referred to as \Upsil, which decays almost exclusively to $\BpBm$ or $\BzBzb$ pairs. 
The maximum production rate for these $\Upsil\to \BB$ events of 20~Hz was achieved at KEK, compared to the multi-hadron non-$\BB$ background rate of about 80~Hz.

$\B$ mesons have very low momenta, $\approx 300$\mev,
and therefore their decay products are distributed almost isotropically in the detector. The BABAR~\cite{Aubert:2001tu,TheBABAR:2013jta} and Belle~\cite{Abashian:2000cg} detectors were designed to cover close to 90\% of the total  
solid angle, thereby enabling the reconstruction of almost all final state particles from decays of the two $B$ mesons, except neutrinos.  

The LHC $pp$ collider operated at total energies of 7 and 8 TeV  from 2008 to 2012.  
In inelastic $pp$ collisions, high energy gluons, the carriers of the strong force, produce pairs of $B$ hadrons (mesons or baryons) along with a large number of other charged and neutral particles, in roughly 1\% of the $pp$ interactions.
The $B$ hadrons are typically produced at small angles to the beam and with high momenta, features that determined the design of the LHCb 
detector~\cite{Alves:2008zz,Dettori:2013xsa}, a single arm forward spectrometer, covering the polar angle range of $3-23$ degrees.  

The high momentum and relatively long $B$ hadron lifetimes result in decay distances of several cm. Very precise measurements of the $pp$ interaction point, combined with the detection of charged particle trajectories from $\B$, $D$ and $\tau$ decay vertices are the very effective method to separate $\B$ decays from background.

\begin{figure*}[btp!]
\centering
\includegraphics[width=0.8\textwidth]{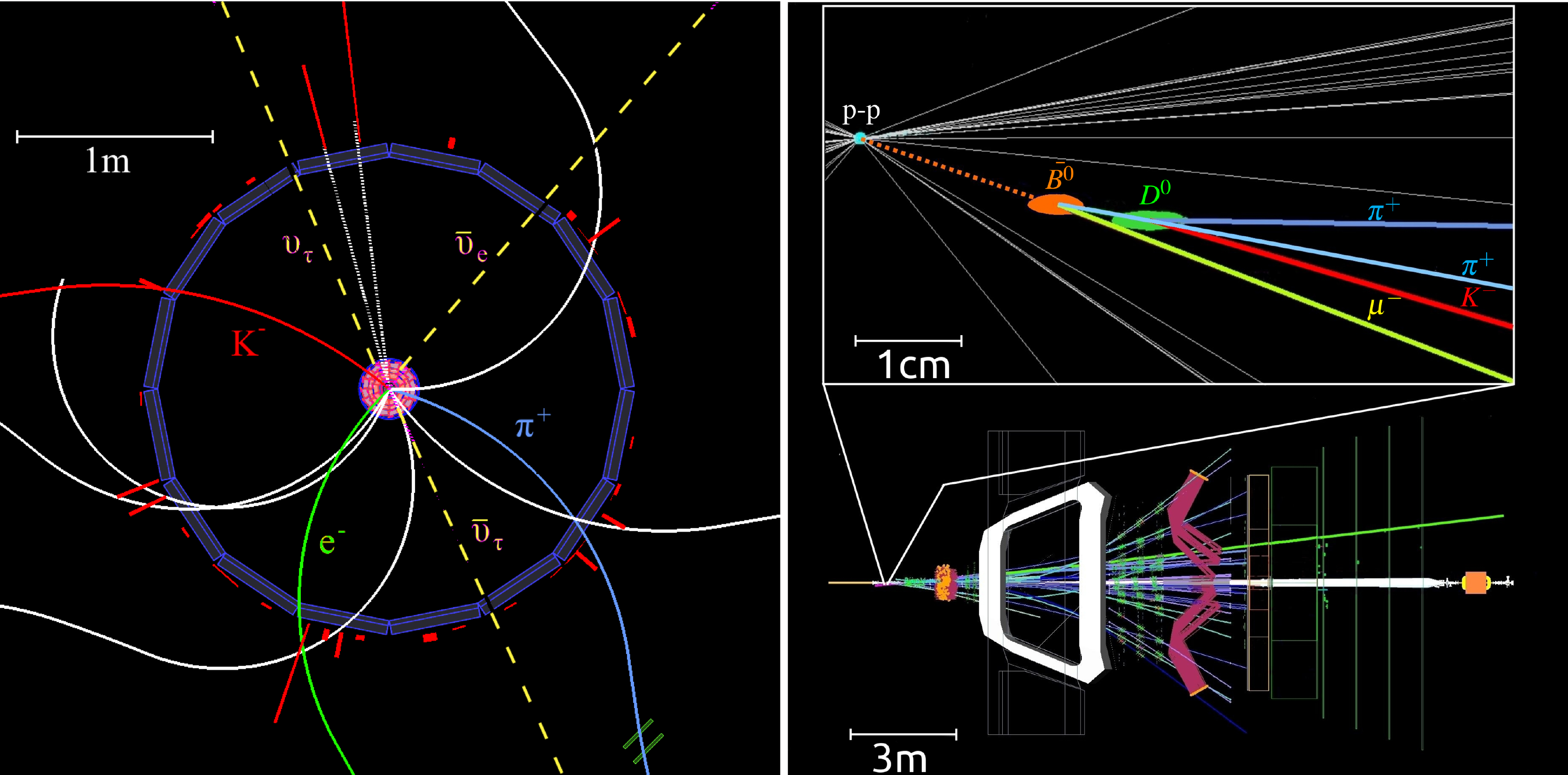}
\vspace{3mm}
\caption{ Belle (a) and LHCb (b) single event displays: Trajectories of charged particles are shown as colored solid lines, energy deposits in the calorimeters are depicted by red bars.
The Belle display is an end view perpendicular to the beam axis with the silicon detector in the center (small orange circle) and the Cherenkov detectors (purple polygon). This is a $\Upsil \to B^+ B^-$ event, with
$B^- \to D^0 \tau^- \bar{\nu_\tau}$, $D^0 \to K^- \pi^+$ and $\tau^- \to e^- \nu_\tau \bar{\nu_e}$, and the 
$B^+$ decaying to five charged  particles (white solid lines) and two photons.  The trajectories of undetected neutrinos are marked as dashed yellow lines.
The LHCb display is a side view with the proton beams indicated as a  white horizontal line with the interaction point far to the left, followed by the dipole magnet (white trapezoid) and the Cherenkov detector (red lines). The area close to the interaction point is enlarged above, showing the tracks of the charged particles produced in the $pp$ interaction, the $\Bz$ path (dotted orange line), and its decay 
$\bar{B^{0}} \to D^{*+} \tau^{-} \bar{\nu}_\tau$ with $D^{*+} \to D^0\pi^+$ and $D^0 \to K^- \pi^+$, plus the $\mu^-$ from the decay of a very short-lived $\tau^{-}$. 
}
\label{fig:events_displays}
\end{figure*}

All three experiments rely on layers of finely segmented silicon strip detectors to locate the beam-beam interaction point and decay vertices of long-lived particles. 
A combination of silicon strip detectors and multiple layers of gaseous detectors measure the trajectories of charged particles deflected in a magnetic field. Devices which sense Cherenkov radiation distinguish charged particles of different masses, and 
arrays of cesium iodide crystals measure the energy of photons and identify electrons at BABAR and Belle.  
Muons are identified as particles penetrating a stack of steel absorbers interleaved with large area gaseous detectors. Examples of reconstructed signal events recorded by the Belle and LHCb experiments are shown in Figure~\ref{fig:events_displays}.

BABAR and Belle exploit the $\BB$ pair production at the $\Upsil$ resonance and have independently developed two sets of algorithms to tag 
$\BB$ events by reconstructing a hadronic or semileptonic decay of one of the two $\B$ mesons, referred to as $B_\mathrm{tag}$.
The hadronic tag algorithms~\cite{Feindt:2011mr,Lees:2013uzd} search for the best match between one of more than a thousand possible decay chains and a subset of all detected particles in the event. 
The efficiency for finding a correctly matched $B_\mathrm{tag}$ is unfortunately small, typically 0.3\%.
The semileptonic tag algorithms relies on a few decays modes with larger branching fractions, resulting in an efficiency of about 1\%.
However, the presence of the neutrino leads to weaker constraints on the $B_\mathrm{tag}$ and more importantly on the signal $\B$ decay.

\section*{Measurements of \texorpdfstring{\BDxtaunu}{B -> D(*)TauNu} Decays}
\label{sec:dxtaunu}

The BABAR and Belle event selection 
required a \Btag, plus a $D$ or $D^*$ meson, and a charged lepton $\ell^-=e^-$ or $ \mu^-$.
Charged and neutral $D$ mesons are reconstructed from combinations of pions and
kaons with invariant masses compatible with the $D$ meson mass.  The higher-mass 
$D^{*0}$ and $D^{*+}$ mesons
are identified by their $D^{*} \rightarrow D\pi$ and $D^{*} \rightarrow D\gamma$ decays.  
Non-$B\bar{B}$ backgrounds and misreconstructed events are greatly suppressed by the \Btag reconstruction. The remaining background is further reduced by multivariate selections.

At LHCb, only decays of $\bar{B}^0$ mesons producing a $D^{*+}$ meson and a $\mu^-$ are selected.
The $D^{*+}$ meson is reconstructed exclusively in $D^{*+} \to D^0 (\to K^- \pi^+) \pi^+ $ decays.  The use of a single decay chain significantly simplifies this analysis and the 
reduced efficiency is compensated by the very large production rate of $B$ mesons at the LHC.  
The bulk of the background is rejected by requiring that all charged particles from the $B$ candidate (and no other tracks)  originate from a common vertex that is significantly separated from the $pp$ collision point.

While for BABAR and Belle the $\B$ momentum is fixed and known precisely, die LHCb
the direction of the $\B$ momentum is inferred from the reconstructed $pp$
collision point and $D^{*+}\mu^-$ vertex, and the magnitude of the $\B$ momentum
is estimated by equating its component parallel to the beam axis to that of the $D^{*+}\mu^-$ combination, rescaled by the
ratio of the $\B$ mass to the measured $D^{*+}\mu^-$ mass.

\begin{figure*}[bt!]
\centering
\includegraphics[width=0.95\textwidth]{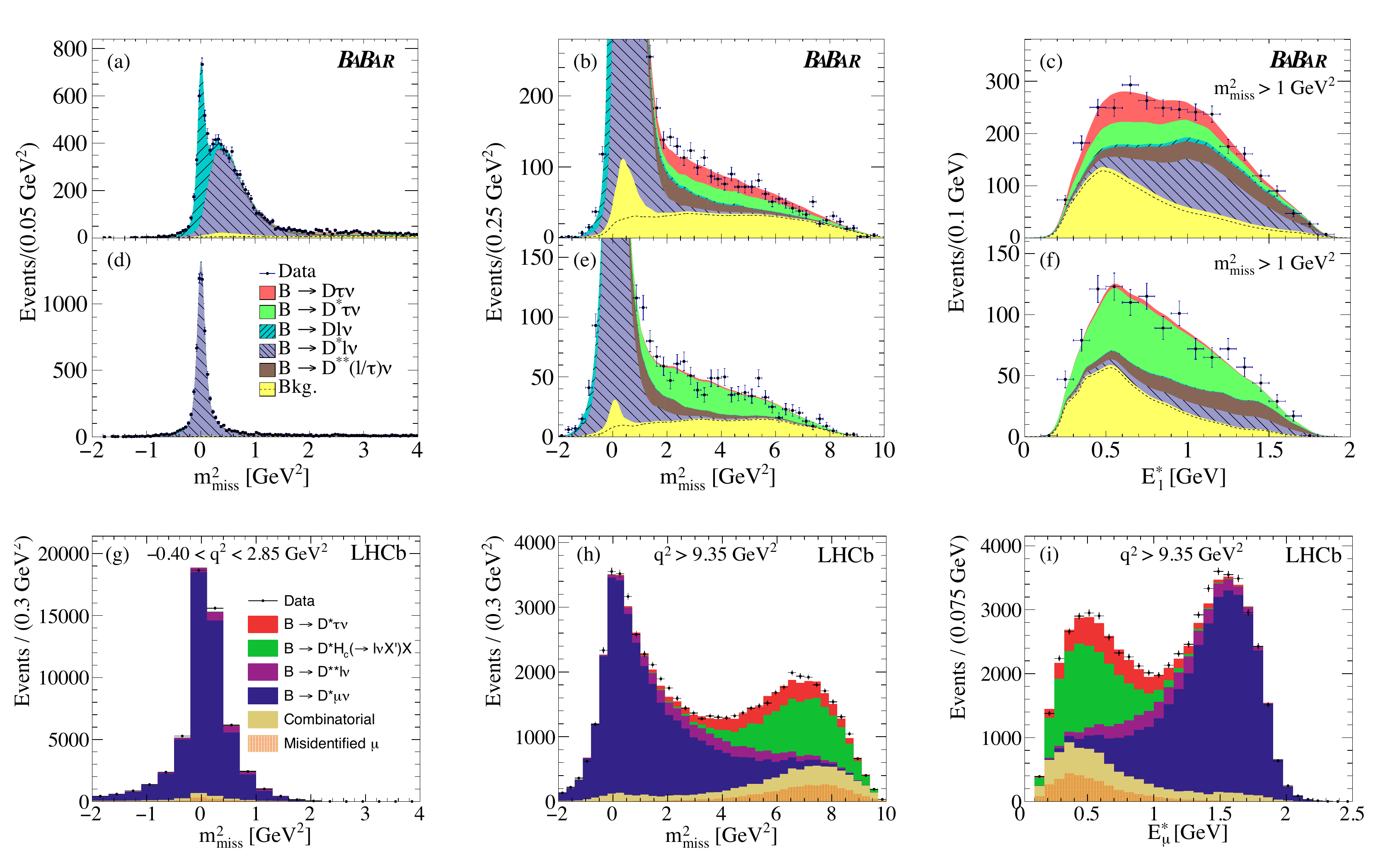}
\caption{ Extraction of the ratios ${\cal R}_D$ and ${\cal R}_{D^*}$ 
by maximum likelihood fits:
Comparison of the projections of the measured \mmiss and \Esl distributions (data points) 
and the fitted distributions of signal and background contributions for the BABAR 
fit~\cite{Lees:2013uzd} to the $D\ell$ samples (a-c) and $D^*\ell$ samples (d-f), as well the
LHCb fit~\cite{Aaij:2015yra} to the $D^{*+}\ell$ sample (g-i).
The \Esl distributions in  (c) and (f) are signal enhanced by the restriction $\mmiss > 1 \gev^2$.  
The  LHCb results are presented for two different $q^2$ intervals, the lowest, which is free of 
$B^0 \to D^{*+}\tau^- \nutb$ decays (g), and the highest where this contribution is large (h,i).}
\label{fig:fit-dxtaunu}
\end{figure*}

The yields for the signal and normalization $\B$ decays, and various background contributions are determined by maximum likelihood fits to the observed data distributions. 
All three experiments rely on three variables:  The invariant mass squared of all undetected particles, $m_\mathrm{miss}^2 =  E_\mathrm{miss}^2 - \vec{p}_\mathrm{miss}^2$, $\Esl$, the energy of the charged lepton in the $B$ rest frame, and $q^2$, the invariant mass squared of the lepton pair.  $E_\mathrm{miss}$ and $\vec{p}_\mathrm{miss}$ refer to the missing energy  and momentum of the $\B$ meson.
BABAR and Belle restrict the data to $q^2>4\text{ GeV}^2$ to enhance the contribution from signal decays. Control samples are used to validate the simulated distributions and constrain the size and kinematic features of the background contributions.

BABAR performs the 2D fit, whereas LHCb divides the $q^2$
range into four intervals, thus performing a fully 3D fit.  
Belle combined two 1D fits, 1) to the \mmiss distribution in the low \mmiss region ($\mmiss < 0.85\text{ GeV}^2$) dominated by the normalization decays, and 2) to a multivariate classifier for data in the high \mmiss region.

Figure~\ref{fig:fit-dxtaunu} shows one-dimensional projections of the data and the
fitted contributions from signal and normalization $\B$ decays, and various backgrounds. The \mmiss distributions for BABAR (and likewise for Belle)
show a narrow peak at zero (Figure~\ref{fig:fit-dxtaunu} a,d), dominated by
normalization decays with a single neutrino, whereas the signal events with three neutrinos extend to about 10 GeV$^2$. For \BDellnu decays, there is a sizable contribution from \BDsellnu decays, for which the pion or
photon from the $D^*\to D\pi$ or $D^*\to D\gamma$ decay was not reconstructed.  For LHCb, the peak at zero is somewhat broader and has a long tail into the signal region (Figure~\ref{fig:fit-dxtaunu} h) because of
uncertainties in the estimation of the \Bsig momentum.  
The \Esl distributions (Figure~\ref{fig:fit-dxtaunu} c,f,i) provide
additional discrimination, since a lepton from a normalization decay has a higher average momentum than a lepton
originating from secondary $\tau^- \to \ell^- \nut \nulb$ decays.

Among the background contributions, semileptonic $B$ decays to the higher mass
$D^{**}$ mesons are of concern, primarily because their branching fractions and form factors are not well known. These $D^{**}$ states decay to a $D$ or $D^{*}$ meson plus additional low energy particles which, if not reconstructed,
have a broader $\mmiss$ distribution. They can be distinguished from signal decays by their \Esl distributions which extend to higher values.  
At LHCb, an important background arises from $B\to D^{(*)}H_c X$ decays, where $H_c$is a charm hadron decaying either leptonically or semileptonically, and $X$ refers to additional low mass hadrons, if present. These decays produce \mmiss and \Esl spectra that are similar to those of signal events (Figure~\ref{fig:fit-dxtaunu} h,i). 

LHCb recently reported a measurement of the ratio ${\cal R}_{D^{*+}}$ using  
$\tau^- \to \pi^- \pi^+ \pi^- (\pi^0)\nutb$ decays~\cite{Aaij:2017deq}.  By requiring a $4 \sigma$ separation of the 3-prong 
vertex from the $B$ decay vertex,  
99\% of the $\Bz \to D^{*+} \pi^- \pi^+ \pi^- (\pi^0)$ background is removed, while 34\% of the signal is retained, and the purity of the signal sample improved by a factor of four compared to the purely leptonic $\tau^-$ decay.

\begin{figure*}[bt!]
\centering
\includegraphics[width=0.95\textwidth]{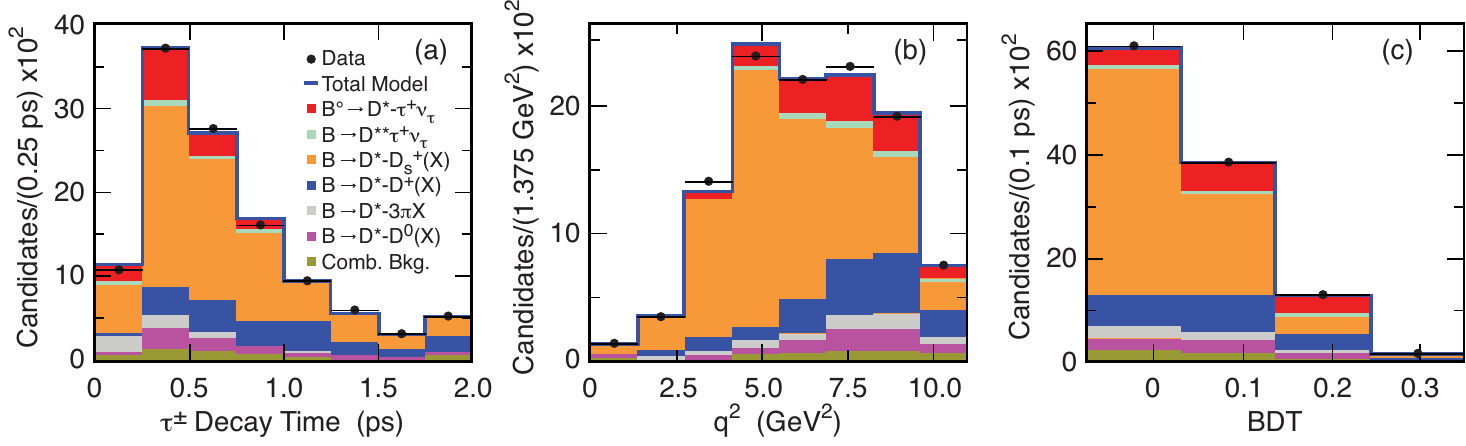}
\caption{LHCb extraction of the $B \to D^{(*)}\tau^- \nutb$ decays with 3-prong $\tau$ vertices by a 3D maximum likelihood fit: 
Projections of distributions of the three variables showing the dominant backgrounds from $B\to D^{(*)}H_c X$ decays~\cite{Aaij:2017deq}. 
}
\label{fig:results-lhcb3pi}
\end{figure*}

The signal yield is extracted via binned fit to a 3D distribution for  $\tau$ decay time (based on the estimated 3-pion momentum), and $q^2$, in 4 bins of the output of the BTD (Boosted Decision Tree) algorithm employed to suppress various backgrounds. Figure~\ref{fig:results-lhcb3pi} shows the projections of the three distributions, resulting in a yield of $(1296 \pm 86)$ signal events.  This rate is normalized to a much larger sample of $\Bzb \to D^{*+} \pi^- \pi^+ \pi^-$ decays.  Correcting for efficiencies and the $\tau$ branching fraction, this translates to
\begin{equation}
\label{eq:K}
{\cal K} \equiv \frac{{\cal B}(\Bzb \to D^{*+} \tau^- \nutb)}
               {{\cal B}(\Bzb \to D^{*+} (3 \pi)^-} 
                     = 1.97 \pm 0.137_{stat}  \pm 0.18_{syst}. \nonumber \
\end{equation}
Taking into account the averages of measurements of the two branching fractions, LHCb quotes 
\begin{eqnarray}
\label{eq:RD*}
{\cal R}_{D^{*+}} &=& {\cal K} \times
\frac {{\cal B}(\Bzb \to D^{*+} (3 \pi)^-} 
      {{\cal B}(\Bzb \to D^{*+} \mu^- \nutb)} \nonumber \\
       &=&  0.291 \pm 0.019_{stat} \pm 0.026_{syst}  \pm 0.013_{ext}. \
\end{eqnarray}
With much larger future data and MC samples this method holds great promise for future analyses exploiting the 3-prong $\tau$ decay vertex. 

\begin{figure}[btp!]
\centering
\includegraphics[width=0.48\textwidth]{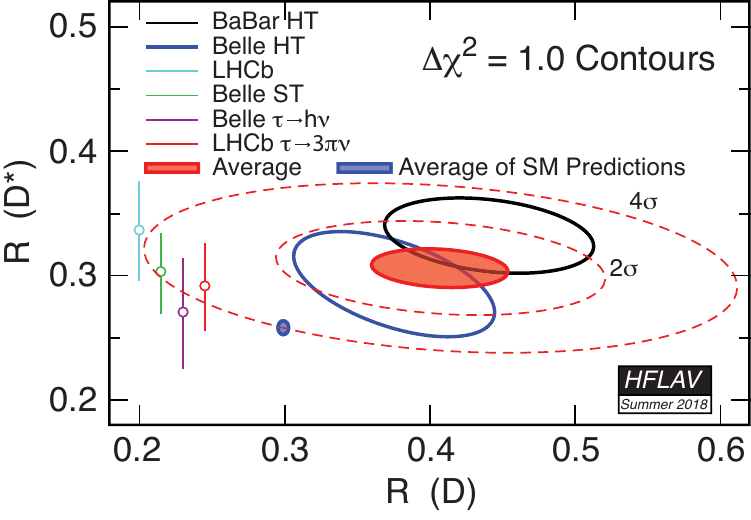}
\caption{Results from 
BABAR~\cite{Lees:2013uzd}, Belle~\cite{Huschle:2015rga,Sato:2016svk}, 
and LHCb~\cite{Aaij:2015yra,Aaij:2017deq}: Their values and 1$\sigma$ contours. The
average calculated by the Heavy Flavor Averaging Group~\cite{HFLAV2018} 
is compared to SM predictions~\cite{Na:2015kha,Fajfer:2012vx,Lattice:2015rga}. 
HT and ST refer to hadronic and semileptonic tag analyses.
}
\label{fig:results-dxtaunu-hflav}
\end{figure}

Figure~\ref{fig:results-dxtaunu-hflav}
shows the measured values for \RD and \RDs by BABAR~\cite{Lees:2013uzd} (including a 
more recent measurement using $\tau^+ \to h^+ \nu _{\tau}$  decays, 
where $h^+$ refers to a $\pi^+$ or $\rho^+$), 
and LHCb~\cite{Aaij:2015yra,Aaij:2017deq}.
The averages of the measurements~\cite{HFLAV2018} are 
\begin{eqnarray}
{\cal R}_D     &=& 0.407 \pm 0.039_{\rm stat} \pm 0.024_{\rm syst}, \\
{\cal R}_{D^*} &=& 0.306 \pm 0.013_{\rm stat} \pm 0.007_{\rm syst}.
\end{eqnarray}
Both values exceed the SM expectations. Taking into account the correlations, the combined difference between
the measured and expected values has a significance of close to four standard deviations.

\section*{Interpretations of Results}
\label{sec:interpretations}

The results presented here have attracted the attention of the physics community and have resulted in several potential explanations of this apparent violation of lepton universality in $\B$ decays involving the $\tau$ lepton.

Among the simplest explanations for these observed rate increases
for decays involving $\tau^-$ 
would be the existence of a new vector boson, $W'^-$, similar to the SM $W^-$ boson, but of  greater mass,
and with couplings of varying strengths to different leptons and quarks.  This could lead to changes
in \RD\ and \RDs, but not in the kinematics of the decays.
However, this option is constrained by searches for $W'^- \to t\bar{b}$
decays~\cite{Chatrchyan:2012gqa,Aad:2014xea} at the LHC collider at CERN, as well as by precision measurements
of $\mu$~\cite{Prieels:2014paa} and $\tau$~\cite{Stahl:2000aq} decays.

Another potentially interesting candidate would be a new type of Higgs boson, a particle of spin 0, similar to the recently discovered neutral Higgs~\cite{Chatrchyan:2012ufa,Aad:2012tfa}, but
electrically charged. 
This charged Higgs ($H^-$) was proposed in
minimal extensions of the SM~\cite{Barger:1989fj}, 
which are part of 
broader theoretical frameworks such as supersymmetry~\cite{Gunion:1984yn}.  The $H^-$ would mediate weak
decays, similar to the $W^-$ (as indicated in Figure 1), but couple differently to leptons of different mass. The $q^2$ and angular distributions would be impacted because of the different spin of the $H^-$.

Another feasible solution might be leptoquarks~\cite{Dorsner:2016wpm}, hypothetical particles with both electric and color (strong) charges that allow transitions from 
quarks to leptons and vice versa, and offer a unified description of three generations of quarks and leptons.
Among the ten different types of leptoquarks, six could contribute to
$B \rightarrow D^{(*)} \tau \nu$ decays~\cite{Freytsis:2015qca}. A diagram of a spin-0 state mediating quark-lepton
transitions is shown in Figure~\ref{fig:feynman}b for the $\B$ decay modes under study.

BABAR and Belle have studied the implications of these hypothetical particles in the context of specific
models~\cite{Lees:2013uzd,Huschle:2015rga}.  The measured values of \RD\ and \RDs\ do not support the simplest
of the two-Higgs doublet models (type II), however, more general Higgs models with appropriate parameter
choices can accommodate the measured ratios~\cite{Datta:2012qk,Crivellin:2012ye,Fajfer:2012jt}.
Some of the leptoquark models could also explain the
measured values of \RD\ and \RDs~\cite{Sakaki:2013bfa,Dumont:2016xpj,Bauer:2015knc}, evading constraints from direct searches
of leptoquarks in $ep$ collisions~\cite{Buchmuller:1986zs} at HERA~\cite{Chekanov:2003af,Collaboration:2011qaa}
and $pp$ collisions at LHC~\cite{ATLAS:2013oea,Khachatryan:2014ura}. 

The kinematics of $B \rightarrow D^{(*)} \tau \nu_{\tau}$ decays should permit further
discrimination of new physics scenarios based on the decay distributions of final state particles.  The
$q^2$ spectrum~\cite{Lees:2013uzd,Huschle:2015rga} and the momentum distributions of the $D^{(*)}$
and electron or muon~\cite{Sato:2016svk} have been examined.  Within the current experimental uncertainties, the observed shapes of these distributions are consistent with SM predictions.

\section*{Conclusions and Outlook}
\label{sec:conclusions}

While the observed enhancements of the semileptonic $B$ meson decay rates involving a $\tau$ lepton relative to the expectations of the SM of electroweak interactions are intriguing,
their significance is not sufficient to unambiguously establish a violation of lepton universality at this time. However, the fact that these unexpected enhancements have been observed by three experiments operating in very different environments deserves further attention.

At present, the measurements are limited by the size of the available
data samples and uncertainties in the reconstruction efficiencies and background estimates.  It is not inconceivable that the experiments have underestimated these uncertainties, or missed a more conventional explanation.  Furthermore, while it is very unlikely, it cannot be totally excluded that the theoretical SM predictions are not as firm as presently assumed. 
Currently, the experimenters are continuing their analysis efforts, refining their methods, enhancing the signal samples by adding additional decay modes, improving the efficiency and selectivity of the tagging algorithms, as well as the Monte Carlo simulations, and scrutinizing all other aspects of the signal extraction. 

At KEK, the $\epem$ collider has undergone major upgrades 
and is expected to enlarge the data sample by close to a factor of 40 
over a period of about ten or more years. 
In parallel, the Belle detector has also been substantially upgraded, and
following commissioning Belle II is expected to be in full operation a year from now.

The much larger event samples should lead to more precise measurements $\B \rightarrow 
D^{(*)} \tau \nu_{\tau}$ decays, based on detailed studies of their kinematics, i.e. $q^2$ and angular distributions, as well as the $\tau$ polarization in $\B \rightarrow D^{*} \tau \nu_{\tau}$ decays. The feasibility of such a measurement has recently been demonstrated~\cite{Hirose:2016wfn}.  
The first measurements of ${\cal R}_{D^{**}}$~\cite{vossen:2018zeg} should lead to a significant reduction of the uncertainties in the estimate of this background for ${\cal R}_{D^{(*)}}$ measurements. Belle analyses will be critical to improved understanding of form factors for various semileptonic $B$and $D$ meson decays.

The unique capabilities of Belle II should allow studies of inclusive semileptonic decays, like $B \to X_u \ell^+ \nut$ ($\ell = e,\mu, \tau $) for both charged and neutral B decays;
$X_u$ refers to the sum of non-charm states $X_u$.
The larger data samples and developments of more refined tagging algorithms will benefit many analyses, in particular 
$\Bm \to \taum \nutb$ decays, should lead to significant reductions in statistical uncertainties and thus allow more stringent tests of SM predictions for these purely leptonic $\B$ meson decays.  

By the end of 2018, the accumulated LHCb data sample is expected to increase by a factor of three.  In the near future, LHCb will complete several important analyses, among them their first measurement of the $\BDtaunu$ decay, which will also improve results for $\BDstaunu$. In parallel, samples with the $\tau^{-} \to \pi^{-} \pi^{+} \pi^{-} \nu_{\tau}$ decay mode 
will be used to improve the signal purity.  

In the longer term future, LHCb is planning to further enhance the trigger selection and data rate capability to record much larger event samples. 

Given the large production rate of $B_s$ and $B_c$ mesons and various $B$ baryons, 
LHCb is planning a broad program to measure their semileptonic branching fractions and form factors and to search for deviation for SM expectations.  For instance, 
$\B_s^0 \to D_s^- \tau^+ \nu_{\tau}$ decays which probe the same interaction as $R_D$.  
LHCb recently observed the decay $\B^+_c \to J/\psi  \tau^+ \nu_{\tau}$ resulting in a final state of 3 muons and three neutrinos,  
and measured the ratio ${\cal R}_{J/\psi} = 0.71 \pm 0.17_{stat} \pm 0.18_{syst}$~\cite{Aaij:2017tyk}. At present, the uncertainties are dominated by the very small signal
and the limited knowledge of the form factor. 
 
Measurements of semileptonic $\Lambda_b$ decays probe different spin structures, specifically the favored $b \to c$ transition 
$\Lambda_b^0 \to \Lambda_c^+ \tau^- \nutb$ and the suppressed $b \to u$ transition 
$\Lambda_b^0 \to p \tau^- \nutb$ measurements that are expected to distinguish between interpretations.
Other $b \to u$ transitions, like 
$B^+ \to \rho^0 \tau^- \nut$ or  $B^+ \to p \pbar \tau^+ \nut$ are also considered.

Independently,
several experiments have examined decay rates and angular distributions for four $\Bp$ decays, 
$\Bp \to K^{(*)+}\mu^+ \mu^-$ and  $\Bp \to K^{(*)+} e^+ e^-$.   
In the framework of the SM, these decays are very rare, since they involve 
$b \to s$ quark transitions.  LHCb~\cite{Aaij:2014ora} published a measurement of the ratio,
\begin{equation}
\label{eq:RK}
{\cal R}_K = \frac {{\cal B}(\Bp \to K^+ \mu^+ \mu^-)}
                        {{\cal B}(\Bp \to K^+ e^+  e^- )} 
                = 0.745 \pm ^{0.090}_{0.074} \pm 0.036 . \
\end{equation}
This results is 2.6 standard deviations smaller than the SM expectation of about 1.0.
Some theoretical new types of interactions could explain this result. For instance, leptoquarks can mediate this decay and result in higher rates for electrons than muons~\cite{Hiller:2014yaa,Becirevic:2016yqi}.
BABAR~\cite{Lees:2012tva}, LHCb~\cite{Aaij:2015oid} and 
Belle~\cite{Wehle:2016yoi} have analyzed angular distributions for the four decay modes and observed general agreement with SM predictions, except for local deviations, the most significant by LHCb at the level of 3.4 standard deviations.  Again, more data are needed to enhance the significance and find possible links to $\B$ decays involving $\tau$ leptons.
 
In conclusion, we can expect much larger event samples from the LHCb and Belle II experiments in the not too distant future.  
These data will be critical to the effort to 
understand whether the tantalizing results obtained to date are an early indication of beyond-the-SM physics processes or the result of larger-than-expected statistical or systematic deviations.
A confirmation of new physics contributions in these decays would shake the foundations of our understanding of matter and trigger an intense program of experimental and theoretical research.


\begin{thebibliography}{10}
\bibitem{charge}
Unless stated otherwise, the inclusion of charged-conjugate states and decay
  modes is implied here and in the following.
\bibitem{helicity}
Helicity refers to the component of the angular momentum of a particle parallel
  to the direction to its momentum.
\end{thebibliography}
\end{document}